\documentclass[prl,floatfix,twocolumn,showpacs]{revtex4}
\usepackage{amsmath}
\usepackage{amsfonts}
\usepackage{mathrsfs}
\usepackage{epsfig}
\usepackage{graphicx}
\usepackage{dcolumn}
\usepackage{amsmath}

\def\be{\begin{equation}}
\def\ee{\end{equation}}

\begin{document}

\title{Are large complex economic systems unstable ?}
\author{Sitabhra Sinha}
\email{sitabhra@imsc.res.in}
\affiliation{%
The Institute of Mathematical Sciences, C. I. T. Campus,
Taramani, Chennai - 600 113, India.}
\begin{abstract}
Although classical economic theory is based on the concept of stable equilibrium,
real economic systems appear to be always out of equilibrium. Indeed, 
they share many of the dynamical features of other complex systems, e.g., ecological 
food-webs. We focus on the relation between increasing complexity of the economic
network and its stability with respect to small perturbations in the dynamical
variables associated with the constituent nodes. Inherent delays and multiple
time-scales suggest that economic systems will be more likely to exhibit instabilities
as their complexity is increased even though the speed at which transactions are
conducted has increased many-fold through technological developments.
Analogous to the birth of nonlinear dynamics from Poincare's work on 
the question of whether the solar system is stable, we suggest that similar
theoretical developments may arise from efforts by econophysicists to understand
the mechanisms by which instabilities arise in the economy. 
\end{abstract}
\pacs{89.65.Gh,05.65.+b,89.75.Da,05.40.Fb}

\maketitle

\newpage

Is the global economic system of the present era inherently unstable? It had long been
thought that the cyclical sequence of inflations and recessions that have buffeted most
national economies throughout the 19th and 20th centuries
are an inevitable result of modern industrial capitalism. However, starting in the 1970s,
economists allied with the influential Chicago school of economics started to promote
the belief that the panacea to all economic ills of the world lay in completely and
unconditionally subscribing to their particular brand of free-market policies. Their
hubris reached its apogee at the beginning of this decade as summed up by the statement
of the Nobel Laureate Robert Lucas at the 2003 annual meeting of the American Economic 
Association that ``the central problem of depression prevention has been solved, 
for all practical purposes''~\cite{Lucas03}. This complacency about the economy's robustness to all
possible perturbations led not only most professional economists, but more importantly,
government bureaucrats and ministers (e.g., Gordon Brown's claims that economic
booms and busts were a thing of the past~\cite{Brown07}) to ignore or downplay 
the seriousness of the present economic and financial crisis at its initial stage. 
As many of the recent books on the onset of the global economic meltdown written 
by Posner and others point out, the mainstream economists and those
whom they advised were blinded by their unquestioning acceptance of the assumptions
of neo-classical theory~\cite{Posner09}.

In response to the rising criticism of traditional economic theory, spearheaded by
physicists working on economic phenomena~\cite{Bouchaud08} as well as non-traditional economists who have collaborated with physicists~\cite{Lux09}, some economists are now trying to put up
a defense that the sudden collapse of markets and banks is not something that can
be predicted by economic theory as this contradicts their basic foundational 
principles of rational expectations
and efficient markets. Thus, according to the conventional economic school of thought,
bubbles cannot exist because any rise in price must reflect all information available
about the underlying asset~\cite{Fama70}. Although detailed analysis of data from 
markets clearly reveals that much of the observed price variation cannot be explained in 
terms of changes in economic fundamentals~\cite{Shiller05}, the unquestioning belief 
in the perfection
of markets has prompted several economists in the past decades to assert that the
famous historical bubbles, such as Tulipomania in 17th century Holland or the South Sea 
Affair of 18th century England, were not episodes of price rise driven by irrational speculation, 
but rather were based on sound economic reasons (see, e.g., Ref.~\cite{Garber90}). 
This complete divorce of theory from 
observations points to the basic malaise of economics. What makes it all the more
worrying is that despite the lack of any empirical verification, such economic theories
have nevertheless been used to guide the policies of national and international agencies
affecting the well-being of billions of human beings. 

In fact, in its desperate effort to become a rigorous science by
adopting, among other things, the formal mathematical framework of
game theory, mainstream economics has become concerned less with
describing reality than with an idealized version of the world. As an
economist recently pointed out, in the overly mathematical formalism
of rational expectations theory, any economic transaction, including
that of a person buying a newspaper from the corner store vendor,
appears to be a complicated chess game between Kenneth Arrow and Paul
Samuelson, two of the most notable post-war economists (quoted in
Ref.~\cite{Sinha10}).  In truth, almost throughout our life, we rarely
go through a complicated optimization process in an effort to
calculate the best course of action. Even if we had access to complete
information about all the options available (which is seldom the
case), the complexity of the computational problem may overwhelm our
decision-making capabilities.  Thus, most often we are satisfied with
choices that seem ``good enough'' to us, rather than the best one
under all possible circumstances. Moreover, our choices may also
reflect non-economic factors such as moral values that are usually not
taken into consideration in mainstream economics.

Given these caveats, it seems that the cherished hypotheses of
efficient markets and rational agents stands on very shaky ground
indeed. The question obviously arises as to whether there are any
alternative foundations that can replace the neo-classical framework.
Behavioral economics, which tries to integrate the areas of
psychology, sociology and economics, is one possible candidate.
Another challenger is from outside the traditional boundaries of
economics, a discipline that has been dubbed
econophysics~\cite{Sinha09,Yakovenko09}. Although physicists have
earlier worked on economic problems occasionally, it was only about a
decade and half ago that a systematic, concerted movement began which
has seen more and more physicists using the tools of their trade to
analyze phenomena occurring in a socio-economic
context~\cite{Farmer05}. This was partly driven by the availability of
large quantities of high-quality data and the means to analyze them
using computationally intensive algorithms. One of the most active
sub-fields within this area is the empirical characterization of
statistical properties of financial markets.  Starting from the work
of Mantegna and Stanley~\cite{Mantegna99}, several important results
are now known about such markets which appear to be {\em universal},
in the sense that they are invariant with respect to the systems being
considered, the time-period under consideration and the type of data
being analyzed. One of the best examples of such universal features of
financial markets is the {\em inverse cubic law} for the distribution
of price (or index) fluctuations~\cite{Gopikrishnan98}.  Not only has
it been observed to hold across several different time-scales and
across different types of stocks (and market indices), but more
surprisingly, it appears to be valid irrespective of the stage of
development of the market~\cite{Sinha07}.

Financial markets have also proved a fertile ground for uncovering the
structure of interactions between the different components of an
economic system. In particular, the transactions between agents buying
and selling different stocks in the market are reflected in the
correlated movements of the prices of different stocks.  Analogous to
the process of inferring the movement of air molecules by watching the
Brownian motion of suspended particles, we can have a coarse-grained
view of the interaction dynamics between individuals in the market by
reconstructing the network of significantly correlated stocks (i.e.,
correlated in terms of their price fluctuations).  Comparison of such
stock interaction networks for different markets has hinted that a
financial market at a later stage of development possesses many more
strongly bound clusters of co-moving stocks that are often from the
same business sector~\cite{Pan07}. As such markets tend to have
identical statistical properties in terms of the distributions of
price or index fluctuations, as well as, other trading indicators,
they differ primarily in the topological structure of the interactions
between their components. Thus, network analysis can provide us with a
window into the process of economic development.

\begin{figure}
\centering
\includegraphics[width=0.98\linewidth]{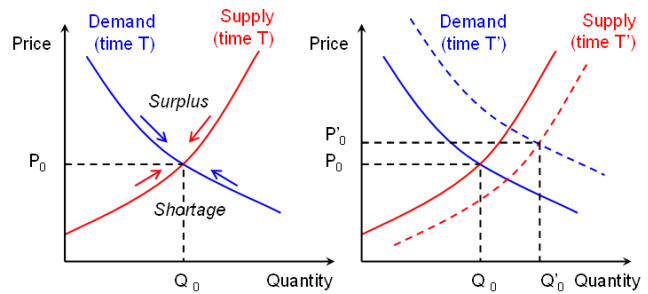}
\caption{{\bf Price mechanism leading to stable equilibrium between supply
and demand according to traditional economic thinking.} (Left) The supply
and demand curves indicate how increasing supply or decreasing demand
can result in falling price or vice versa. If the available supply of a certain
good in the market at any given time is less than its demand for it among 
consumers, its price will go up. The perceived shortage will stimulate increase
in production that will result in an enhanced supply. However, if supply increases
beyond the point where it just balances the demand at that time, there will be
unsold stock remaining which will eventually push the price down. This in turn will
result in a decrease in the production. Thus, a negative feedback control mechanism
governed by price will move demand and supply along their respective curves
to the mutual point of intersection, where the quantity available ($Q_0$) at the
equilibrium price $P_0$ is such that supply exactly equals demand. (Right) As
the demand and supply of a product changes over time due to various different
factors, the supply and demand curves may shift on the
quantity-price space. As a result, the new equilibrium will be at a different
price ($P_0^{\prime}$) and quantity ($Q_0^{\prime}$). Until the curves shift again,
this equilibrium will be stable, i.e., any perturbation in demand or supply will
quickly decay and the system  will return to the equilibrium.
}
\label{fig:price_eqlbm}
\end{figure}
When we broaden our problem from the relatively restricted context of financial markets
to general economic phenomena, the role played by networks of interactions become even more
intriguing. Traditionally, economics has been concerned primarily with equilibria.
Fig.~\ref{fig:price_eqlbm} shows that the price mechanism was perceived by economists
to introduce a negative feedback between perturbations in demand and supply, so that
the system quickly settles to the equilibrium where supply exactly equals demand.
Much of the pioneering work of Samuelson~\cite{Samuelson47}, Arrow~\cite{Arrow58} and 
others (for a review see Ref.~\cite{Negishi62}) had been involved with
demonstrating that such equilibria can be stable, subject to 
several restrictive conditions. However, the occurrence of complex networks of 
interactions in reality bring new dynamical issues to fore. Most notably, we are faced
with the question: do complex economic networks give rise to instabilities ?
Given that most economic systems at present are composed of numerous strongly connected  
components, will periodic and chaotic behavior be the norm for such systems rather
than static equilibrium solutions ?

This question has, of course, been asked earlier in different contexts. In ecology,
it has given rise to the long-standing stability-diversity debate~\cite{May73}.
In the network framework, the ecosystem can be thought of as a network of species,
each of the nodes being associated with a variable that corresponds to the population
of the species it represents. The stability of the ecosystem is then defined by the
rate at which small perturbations to the populations of various species decay with time.
If the disturbance instead grows and gradually propagates through the system affecting
other nodes, the equilibrium is clearly unstable. Prior to the pioneering work of
May in the 1970s, it was thought that increasing complexity of an ecosystem, either
in terms of a rise in the total number of species or the density and strength of their
connections, results in enhanced stability of the ecosystem. This belief was based
on empirical observations that more diverse food-webs (e.g., in the wild) showed less violent
fluctuations in population density than simpler communities (such as in fields
under monoculture) and were less likely to suffer species extinctions. It has also been
seen that tropical forests, which generally tend to be more diverse than sub-tropical
ones, are also more resistant to invasion by foreign species~\cite{Elton58}. It was
therefore nothing short of a shock to the field when in 1972, Robert May showed in the
very brief article {\em Will a large complex system be stable?}~\cite{May72} 
using linear stability arguments that as complexity increases, a randomly connected
network would tend to become more and more unstable. 

The surprising demonstration that a system which has many elements and/or dense connections
between its elements is actually more likely to suffer potentially damaging large
fluctuations initiated by small perturbations immediately led to a large body of 
work on this problem (see Ref.~\cite{McCann00} for a review). The two major objections
to May's results were (a) it uses linear stability analysis and that (b) it assumed random
organization of the interaction structure. However, more recent work which consider 
systems with different types of population dynamics in the nodes, including 
periodic limit-cycles and chaotic attractors~\cite{Sinha05b,Sinha06}, as well as,
networks having realistic features such as clustered small-world 
property~\cite{Sinha05a} and scale-free degree distribution~\cite{Sinha05c},
have shown the result of increasing instability of complex networks to be 
extremely robust. While large complex networks can still arise as a result
of gradual evolution~\cite{Wilmers02}, it is almost inevitable that such
systems will be frequently subject to large fluctuations and extinctions.

The relevance of this body of work to understanding the dynamics of economic
systems has been highlighted in the wake of the recent banking crisis when a
series of defaults, following each other in a cascading process, led to the 
collapse of several major financial institutions. In fact, May and two other theoretical ecologists have written
an article entitled {\em Ecology for bankers}~\cite{May08} to point out the
strong parallels between understanding collapse in economic and ecological
networks. Recent empirical determination of networks occurring in the financial
context, such as that of inter-bank payment flows between banks through the 
{\em Fedwire} real-time settlement service run by the US Federal Reserve, has
now made it possible to analyze the process by which cascades of failure events
can occur in such systems~\cite{Soramaki07}. Analogous to ecological systems, where population
fluctuations of a single species can trigger diverging deviations from the
equilibrium in the populations of other species, congestion in settling the
payment of one bank can cause other pending settlements to accumulate rapidly
setting up the stage for a potential major failure event. It is intriguing
that it is the very complexity of the network that has made it susceptible
to such network propagated effects of local deviations which makes global or
network-wide failure even more likely.
As the world banking system becomes more and more connected, it may be 
very valuable to understand how the topology of interactions can
affect the robustness of the network.

The economic relevance of the network stability arguments used in the ecological
context can be illustrated
from the following toy example. Consider a model financial market comprising
$N$ agents where
each agent can either buy or sell at a given time instant. 
This tendency can be quantitatively measured by the
probability to buy, $p$, and its complement, the probability to sell, $1-p$.
For the market to be in equilibrium, the demand should equal supply, so that
as many agents are likely to buy as to sell, i.e., $p = 0.5$. Let us in addition
consider that agents are influenced in their decision to buy or sell by the
actions of other agents with whom they have interactions. In general, we can 
consider that out of all possible pairwise interactions between agents only a fraction
$C$ are actually realized. In other words, the inter-agent
connections are characterized by the matrix of link 
strengths {\bf J}=$\{J_{ij}\}$ (where $i,j=1, ..., N$ label the agents) 
with a fraction $C$ of non-zero entries. 
If $J_{ij}>0$, it implies that an action of agent 
$j$ (buying or selling) is likely to influence agent $i$ to act in the 
same manner, whereas $J_{ij}<0$ suggests that the action of $i$ will be contrary
to that of $j$. Thus, the time-evolution of the probability for agent $i$ to buy
can be described by the following linearized equation close to the equilibrium
$p_i = 0.5 (i=1,\ldots,N)$:
\begin{equation}
\frac{dp_i}{dt}=\epsilon_i (0.5 - p_i) + \Sigma_j J_{ij} (0.5-p_j),
\label{eq1}
\end{equation}
where $\epsilon_i$ is the rate of converge of an isolated node
to its equilibrium state of equal probability for buying or selling. Without much
loss of generality we can consider $\epsilon_i = 1$ by appropriate choice of 
time units for the dynamics. If in addition, we consider that for simplicity the
interactions are assigned randomly from a Gaussian distribution with mean $0$ and
variance $\sigma^2$, then the largest eigenvalue of the corresponding Jacobian 
matrix {\bf J} evaluated around the equilibrium is $\lambda_{max} 
= \sqrt{NC\sigma^2-1}$. For system parameters such that $NC\sigma^2 > 1$, an
initially small perturbation will gradually grow with time and drive the
system away from its equilibrium state.
Thus, even though the equilibrium $p=0.5$ is stable for 
individual nodes in isolation, it may become unstable under
certain conditions when interactions between the agents are introduced. 
Note that the argument can be easily generalized to the case where 
the distribution from which $J_{ij}$ is chosen has a non-zero mean.

\begin{figure}
\centering
\includegraphics[width=0.98\linewidth]{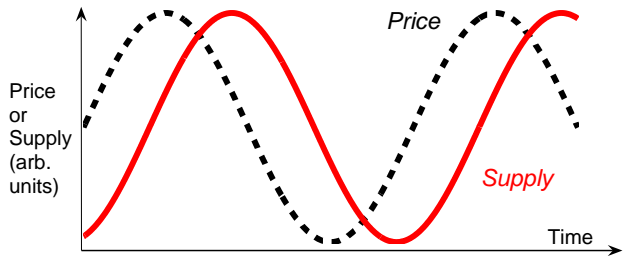}
\caption{{\bf Delay in market response can result in persistent price
oscillations.} Ideally the price mechanism should result in a
transient increase (decrease) in demand to be immediately matched by a
corresponding increase (decrease) in supply. However, in reality there
is delay in the information about the rise or fall in demand reaching
the producer; moreover, at the production end it may take time to
respond to the increasing demand owing to inherent delays in the
production system. Thus, the supply may always lag behind the price in
a manner that produces oscillations: as price rises, supply initially
remains low before finally increasing, by which time demand has fallen
due to the high price which (in association with the increased supply)
brings the price down. Supply continues to rise for some more time
before starting to decrease. When it falls much lower than the
demand, the price starts rising again which starts the whole cycle
anew. Thus, if the demand fluctuates at a time-scale that is shorter
than the delay involved in adjusting the production process to respond
to variations in demand, the price may evolve in a periodic or even a
chaotic manner.
}
\label{fig:price_osc}
\end{figure}
Another problem associated with the classical concept of economic equilibrium
is the process by which the system approaches it. Walras, in his original formulation
of how prices achieve their equilibrium value had envisioned the {\em t\^{a}tonnement}
process by which a market-maker takes in buy/sell bids from all agents in market
and gradually adjusts price until demand equals supply. Formally, it resembles
an iterative convergence procedure for determining the fixed-point solution of a
set of dynamical equations. However, as we know from the developments in nonlinear dynamics
over the past few decades, such operations on even simple non-linear systems
(e.g., the logistic equation) can result in periodic cycles or even chaos~\cite{May76}.
It is therefore not surprising to consider a situation in which the price mechanism
can actually result in supply and demand to be forever out of step each other even though
each is trying to respond to changes in the other. A simple situation in which
such a scenario can occur is shown in Fig.~\ref{fig:price_osc}, where a delay in
the response of the supply to the changes in price through variations in demand
can cause persistent oscillations.

As the principal reason for the instability appears to be the delay, one can
argue that by increasing the speed of information propagation
it should be possible to stabilize the equilibrium. However, we seem to have
witnessed exactly the reverse with markets becoming more volatile as improvements
in communication enable economic transactions to be conducted faster and faster.
As a history of financial manias and panics points out, ``there is little historical
evidence to suggest that improvements in communications create docile financial
markets \ldots''~\cite{Chancellor99}.
A possible answer to this apparent paradox lies in the fact that in
any realistic economic situation, information about fluctuations in the demand
may require to be relayed through several intermediaries before it reaches
the supplier. In other situations, the market may be segmented into several
communities of agents, with significantly more interactions occurring
between agents within the same as opposed to different communities.
These features can introduce several levels of delays in the market, resulting
in a multiple time-scale problem~\cite{Pan09}. Thus, increasing the speed of transactions,
while ostensibly allowing faster communication at the global scale can
disrupt the dynamical separation between process operating at different
time-scales. This can prevent sub-systems from converging to their respective
equilibria before subjecting them to new perturbations, thereby always keeping
the system out of equilibrium.

Therefore, we see that far from conforming to the neo-classical ideal of a stable
equilibrium, the dynamics of the economic system is likely to be always far
from equilibrium. In analogy with the question asked about ecological and other
systems with many diverse interacting components, we can ask whether a sufficiently 
complex economy is bound to exhibit instabilities. After all, just like the
neo-classical economists, natural scientists also at one time believed in the
clockwork nature of the physical world (which in turn influenced the
the English philosopher, Thomas Hobbes, to seek the laws for social organization).
However, Poincare's work on the question of whether the solar system is stable
showed the inherent problems with such a viewpoint and eventually paved the
way for the later developments of chaos theory. Possibly we are at the brink of
a similar theoretical breakthrough in econophysics, one that does not strive to
re-interpret (or even ignore) empirical data
so as to conform to a theorist's expectations but one which describes the 
mechanisms by which economic
systems actually evolve over time. It may turn out that, far from failures of the
market that need to be avoided, crashes and depressions may be the necessary
ingredients of future developments, as has been suggested by Schumpeter in
his theory of {\em creative destruction}~\cite{Schumpeter75}.
However, most importantly, we should not forget that economic
phenomena form
just one aspect of the entire set of processes that make up the human social
organization. Econophysics has to ultimately strive to be a theory for the
entire spectrum of human social behavior. As Keynes, one of the greatest 
economists, had once said 
``do not let us overestimate the importance of the economic problem, or sacrifice
to its supposed necessities other matters of greater and more permanent significance''
\cite{Keynes31}.

\end{document}